\documentclass[prb,onecolumn,showpacs,preprintnumbers,amsmath,amssymb,assume]{revtex4}
\usepackage{graphicx}				% Include figure files (!) No mixing .eps and .pdf figures
\usepackage{dcolumn}			   % Align table columns on decimal point
\usepackage{bm}						  % Bold math
\usepackage{cases}
\usepackage{color}
\begin{document}

\title{NMO-velocity surfaces and Dix-type formulae \\ in anisotropic heterogeneous media}  
\author{Vladimir Grechka$^{1}$ and Ilya Tsvankin$^{2}$}
\affiliation{$^{1}$Shell International Exploration and Production, currently at Marathon Oil Company \\
$^{2}$Colorado School of Mines}
\date{\today}

\begin{abstract}
	
Reflection moveout of pure modes recorded on conventional-length spreads is described by a normal-moveout (NMO) velocity that depends on the orientation of the common-midpoint (CMP) line. Here, we introduce the concept of NMO-velocity surfaces, obtained by plotting the NMO velocity as the radius-vector along all possible directions in 3-D space, and use it to develop Dix-type averaging and differentiation algorithms in anisotropic heterogeneous media.

The intersection of the NMO-velocity surface with the horizontal plane represents the NMO ellipse that can be estimated from wide-azimuth reflection data. We demonstrate that the NMO ellipse and conventional-spread moveout as a whole can be modeled by Dix-type averaging of specifically oriented cross-sections of the NMO-velocity surfaces along the zero-offset reflection raypath. This formalism is particularly simple to implement for a stack of homogeneous anisotropic layers separated by plane dipping boundaries. Since our method involves computing just a single (zero-offset) ray for a given reflection event, it can be efficiently used in anisotropic stacking-velocity tomography. 

Application of the Dix-type averaging to layered transversely isotropic media with a vertical symmetry axis (VTI) shows that the presence of dipping interfaces above the reflector makes the $P$-wave NMO ellipse dependent on the vertical velocity and anisotropic coefficients $\epsilon$ and $\delta$. In contrast, $P$-wave moveout in VTI models with a horizontally layered overburden is fully controlled by the NMO velocity of horizontal events and the Alkhalifah-Tsvankin coefficient $\eta \approx \epsilon - \delta$. Hence, in some laterally heterogeneous, layered VTI models $P$-wave reflection data may provide enough information for anisotropic depth processing. 
	
\end{abstract}
   % 81.05.Xj for anisotropy, 91.30.-f for seismology
\maketitle

\section{Introduction} \label{sec:intro}

NMO velocity estimated from reflection traveltimes recorded in common-midpoint geometry provides valuable information about the velocity field and anisotropic parameters of the subsurface\cite{Tsvankin2001}. Although the relationship between the measured moveout velocity and elastic parameters becomes rather complicated if the model is heterogeneous and anisotropic, the azimuthal dependence of NMO velocity has a simple explicit form.
Grechka and Tsvankin\cite{GrechkaTsvankin1998} examined pure-mode reflection 
traveltimes recorded at a fixed CMP location along different azimuths $\alpha$ in the horizontal plane and showed that the NMO velocity $V_{\rm nmo}(\alpha)$ typically varies as an {\em ellipse}$\,$ in the horizontal plane. [$V_{\rm nmo}(\alpha)$ may have a different form only if CMP traveltime decreases (i.e., reverse moveout) with offset in one or more directions.] In the special case of a homogeneous isotropic layer, the NMO ellipse was first obtained by Levin\cite{Levin1971}.

The orientation and semi-axes of the NMO ellipse depend on the spatial derivatives of the slowness vector at the CMP location. The simplicity and generality of this result arise because NMO velocity governs the wavefront curvature at zero offset\cite{Shah1973}; therefore, its azimuthal variation has to be a quadratic function in the spatial coordinates. Grechka, Theophanis and Tsvankin\cite{GrechkaTheophanisTsvankin1999} extended the equation of the NMO ellipse to mode-converted waves in horizontally layered anisotropic models with a horizontal symmetry plane in each layer.

The elliptical azimuthal dependence of the NMO-velocity function was used to develop efficient algorithms for azimuthal stacking-velocity analysis and moveout correction in wide-azimuth 3-D surveys\cite{Corriganetal1996, GrechkaTsvankin1999}. Even more importantly, the equation of the NMO ellipse provides a foundation for moveout inversion in arbitrary anisotropic media. For models with a horizontally layered overburden above a dipping reflector, the NMO ellipse at the surface represents a Dix-type average of the interval NMO ellipses\cite{GrechkaTsvankinCohen1999}. This equation, generalizing the classical Dix\cite{Dix1955} result and its extensions for isotropic media\cite{Shah1973, HubralKrey1980}, can be used to reconstruct interval NMO ellipses from surface data and then invert them for the anisotropic parameters. The parameter-estimation methodology based on this approach was successfully implemented for several common anisotropic models including orthorhombic media\cite{GrechkaTsvankin1999ORT} and transverse isotropy with a vertical\cite{GrechkaTsvankin1998}, horizontal\cite{Contrerasetal1999HTI} and tilted\cite{GrechkaTsvankin2000TTI} symmetry axis. 

The papers on parameter estimation listed above, however, consider only laterally homogeneous models above the reflector. Although Grechka, Tsvankin and Cohen\cite{GrechkaTsvankinCohen1999} outlined an approach for computing NMO ellipses in arbitrary heterogeneous media, their algorithm is purely numerical and is difficult to apply in interval parameter estimation. The correction of NMO ellipses for lateral velocity variation introduced by Grechka and Tsvankin\cite{GrechkaTsvankin1999} is restricted to horizontal layers with a horizontal symmetry plane.

Here, we relax the assumption of Grechka and Tsvankin\cite{GrechkaTsvankin1998} that
the CMP line belongs to the horizontal plane and examine the NMO velocity 
measured along an arbitrary direction ${\cal L}$
in 3-D space. (One can imagine, for instance, recording reflection 
arrivals along an oblique or vertical borehole.) If the vectors $V_{\rm nmo}({\cal L})$ are plotted from the CMP location, their ends form the NMO-velocity surface, while the NMO ellipse is the 
intersection of this surface with the horizontal plane. 

Even though under normal circumstances we cannot count on measuring $V_{\rm nmo}$ along many different directions in space, this new theory naturally 
leads to a concise Dix-type representation of NMO ellipses in heterogeneous anisotropic media. As an example, we construct Dix-type formulae for a stack of homogeneous anisotropic layers separated by plane dipping interfaces. In the practically important case of non-elliptical VTI media, the influence of dipping interfaces above the reflector may make the $P$-wave NMO ellipses measured at the surface dependent on the interval vertical velocities and Thomsen's coefficients $\epsilon$ and $\delta$, thus affording opportunities for reconstructing the model in depth using $P$-wave moveout data. 

\section{NMO-velocity surfaces in heterogeneous anisotropic media}

\subsection{General formulation} 

We consider the NMO velocity of a pure-mode reflected wave
that was recorded along an arbitrary oriented CMP line ${\cal L}$ in heterogeneous
anisotropic media. It is assumed that the traveltime of the selected reflection event is uniquely defined for each (moderate compared to the reflector depth) source-receiver offset. If the traveltime becomes multi-valued, as in the vicinity of shear-wave cusps, the moveout function usually requires a more elaborate approximation than the hyperbolic equation parameterized by NMO velocity.

The exact function $V_{\rm nmo}({\cal L})$ is derived in
Appendix A as [see equation~(\ref{eq20a})]
\begin{equation}
  V_{\rm nmo}^{-2}({\cal L}) = 
  {\cal L} \, {\bf U} \, {\cal L}^{\bf T} \, ,
  \label{eq01}
\end{equation}
where ${\cal L} = [{\cal L}_1, \, {\cal L}_2, \, {\cal L}_3]$ is a unit row vector, ${\cal L}^{\bf T}$ is a unit column vector, and ${\bf U}$ is a 
$3 \times 3$ symmetric matrix  with the elements  
\begin{equation}
  U_{km} = \tau_0 \, \frac{\partial^2 \tau({\bf x})}
                                   {\partial x_k \, \partial x_m} \equiv
           \tau_0 \, \frac{\partial p_k({\bf x})}{\partial x_m} \, , 
           \qquad (k,m = 1,2,3) \, . 
  \label{eq02}
\end{equation}
Here $\tau_0$ is the one-way traveltime from the zero-offset reflection point to the CMP location, and
$p_k({\bf x})$ are the components of the slowness vector
${\bf p({\bf x})} = [p_1({\bf x}), p_2({\bf x}), p_3({\bf x})]$ corresponding to rays excited at the zero-offset reflection point and recorded at location ${\bf x}$. The derivatives in equation~(\ref{eq02}) are evaluated at the common midpoint.
 
Azimuthally dependent NMO velocity in the horizontal plane (usually an ellipse) described by Grechka and Tsvankin\cite{GrechkaTsvankin1998} can be viewed as the intersection
of the NMO-velocity surface ${\bf U}$ with the horizontal plane. Substituting a horizontal unit vector ${\cal L}^{\rm hor} = [\cos \alpha, \, \sin \alpha, \, 0]$ into the general expression~(\ref{eq01}) yields the NMO ellipse as a function of the azimuth $\alpha$,
\begin{equation}
  V_{\rm nmo}^{-2}(\alpha) = 
  U_{11} \cos^2 \alpha + 2 \, U_{12} \sin \alpha \cos \alpha +
  U_{22} \sin^2 \alpha \,  .
  \label{eq05}
\end{equation}
The $2 \times 2$ matrix ${\bf W}$ introduced by Grechka and Tsvankin\cite{GrechkaTsvankin1998} to define the NMO ellipse~(\ref{eq05}) coincides with the upper left submatrix of ${\bf U}$:

\begin{equation}
W_{ij} = \tau_0 \, \frac{\partial p_i({\bf x})}{\partial x_j} = U_{ij} \, , \qquad (i,j = 1,2) \, . 
  \label{eq06}
\end{equation}

\subsection{Possible shapes of NMO-velocity surfaces}

Equation~(\ref{eq01}) indicates that the function $V_{\rm nmo}({\cal L})$ defines 
a {\em centered quadratic}$\,$ surface in 3-D space. The shape of this surface is determined by the eigenvalues of the matrix ${\bf U}$, which have to be real because ${\bf U}$ is real and symmetric. Using equation~(\ref{eq06}), ${\bf U}$ can be written in the form
\begin{equation}
  {\bf U} = \left( \begin{array}{ccc}
                   W_{11} & W_{12} & U_{13} \\ 
                   W_{12} & W_{22} & U_{23} \\ 
                   U_{13} & U_{23} & U_{33} 
                   \end{array} \right) .
  \label{eq07}
\end{equation}
Equation~(\ref{eq07}) allows us to use the known properties of the matrix ${\bf W}$\cite{GrechkaTsvankin1998, GrechkaTsvankinCohen1999} to make several 
important observations about the matrix ${\bf U}$. First, if the NMO ellipse 
${\bf W}$ has been found from moveout data, we need to determine only three quantities $U_{k3}$ to reconstruct the whole NMO-velocity surface ${\bf U}$. Below, we show that the elements $U_{k3}$ can be computed by differentiating the Christoffel equation at the CMP location. Second, since for homogeneous anisotropic 
media the elements $W_{ij}$ can be obtained in an explicit form\cite{GrechkaTsvankinCohen1999}, the matrix ${\bf U}$ as a whole 
can be found explicitly as well. Third, the cross-section of the NMO-velocity
surface ${\bf U}$ along the horizontal plane is elliptical because the matrix 
${\bf W}$ typically represents an ellipse in the horizontal plane\cite{GrechkaTsvankin1998}.

\begin{figure}
	\centerline{\begin{tabular}{c}
			\includegraphics[width=12.5cm]{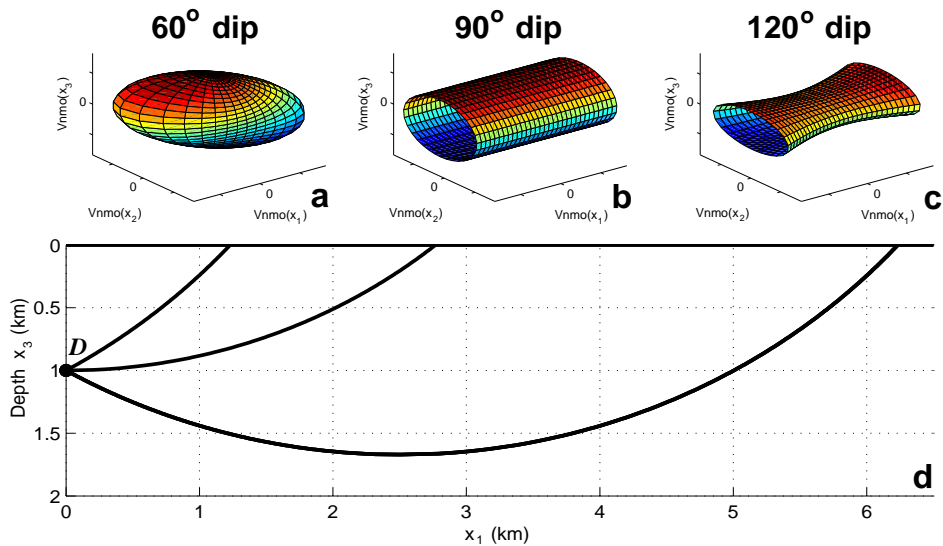} \\
	\end{tabular}}
\caption{NMO surfaces (a)~--~(c) and trajectories of the zero-offset rays (d) in  an isotropic medium with a constant vertical-velocity gradient. Reflector dip is  (a) 60$^\circ$, (b) 90$^\circ$ and (c) 120$^\circ$. The zero-offset reflection point $D$ is located at a depth of 1~km; the parameters of the velocity function $V(x_3) = V_0 + \nu x_3$ are $V_0 = 2.0$~km/s and $\nu = 0.6$~s$^{-1}$.
}
\label{fig01}
\end{figure}

There are only three distinct types of quadratic surfaces which have elliptical cross-sections symmetric with respect to the CMP: ellipsoids, elliptical cylinders and one-sheeted hyperboloids, as shown in Figure~\ref{fig01}. (Note that a hyperboloid and a cylinder may also have a non-elliptical intersection with the horizontal plane, as illustrated below.) 
Since the NMO-velocity surface is quadratic, it may have other shapes, such as those of a two-sheeted hyperboloid, an imaginary elliptical cylinder, 
a hyperbolic cylinder, etc. However, this may happen only if the matrix 
${\bf U}$ has at least two non-positive eigenvalues, and reflection traveltime does not increase with offset in two or more directions in space. Although such cases are not prohibited by the theory, their occurrence is expected to be rare.  

Note that if the NMO-velocity surface has the form of a cylinder, the NMO velocity along the axis of the cylinder is infinite,
which implies that the traveltime in this direction does not change with offset. A numerical example below shows that both elliptical cylinders and one-sheeted hyperboloids can be encountered in realistic subsurface models; a more comprehensive discussion of cylindrical NMO-velocity surfaces is presented below.

\subsection{Example of NMO-velocity surfaces}

To investigate the shape of NMO-velocity surfaces for typical seismological models, we computed the matrix ${\bf U}$ for an isotropic medium
with a constant vertical-velocity gradient [the velocity function is defined as $V(x_3) = V_0 + \nu x_3$]. For this
model, the one-way traveltime $\tau ({\bf x})$ from the origin of the coordinate system to point ${\bf x} = [x_1, x_2, x_3]$ can be found analytically\cite{Slotnick1959},
\begin{equation}
\tau({\bf x}) = \frac{1}{\nu} \, 
  \cosh^{-1} \left[ 1 + \frac{\nu^2 \, (x_1^2 + x_2^2 + x_3^2)}
                             {2 \, V_0 \, (V_0 + \nu x_3)} \right] \, . 
  \label{eq077}
\end{equation}
Using equations~(\ref{eq02}) and~(\ref{eq077}), we derived explicit  
expressions for the matrix ${\bf U}$ in terms of $V_0$, $\nu$, and the depth and dip of the reflector.

Figure~\ref{fig01} displays the NMO surfaces, along with computed ray trajectories
(circular arcs), for reflectors beneath this constant-gradient isotropic medium. Depending on reflector dip, the NMO-velocity surface can take any of the three shapes discussed above (an ellipsoid, a cylinder and a hyperboloid). Numerical tests for other isotropic models, where velocity monotonically increases with depth, indicate that NMO-velocity ellipsoids correspond to reflector dips below 90$^\circ$, cylinders to vertical reflectors, and hyperboloids to dips exceeding 90$^\circ$ (overhangs).

\section{Computation of NMO-velocity surfaces}

\subsection{Heterogeneous arbitrary anisotropic media}

The matrix ${\bf U}$ contains six quantities [equation~(\ref{eq07})] including three components of the matrix ${\bf W}$ responsible for the NMO ellipse. Computation of ${\bf W}$ was described by Grechka, Tsvankin and Cohen\cite{GrechkaTsvankinCohen1999} and is further discussed below. The remaining elements $U_{3k}$ of the matrix ${\bf U}$ depend on the spatial derivatives of the vertical slowness component $p_3$, which can be obtained by combining $W_{ij}$ with the solution of the Christoffel equation at the common midpoint. Indeed, the slownesses at each spatial location ${\bf x}$ are related to each other by the Christoffel equation, which can be written as
\begin{equation}
  F({\bf p}, {\bf x}) = 0 \, , 
  \label{eq08}
\end{equation}
where ${\bf p}$ is the slowness vector at the spatial location ${\bf x}$ (e.g.,
Grechka, Tsvankin and Cohen, 1999). Equation~(\ref{eq08}) contains a separate contribution of the coordinates ${\bf x}$ because of the spatial dependence of the stiffness coefficients $c_{ij}$ in heterogeneous media. At a fixed location ${\bf x}$, equation~(\ref{eq08}) allows us to express the vertical slowness $p_3$ through the horizontal slownesses $p_1$ and $p_2$. Since the elements $U_{3k}$ depend on $p_3$, they can be found as functions of the derivatives $\partial p_1/ \partial x_k$ and $\partial p_2/ \partial x_k$ that determine the NMO ellipse ${\bf W}$. A complete derivation given in Appendix B leads to the following expression for the matrix ${\bf U}$ [equation~(\ref{eq09b})]:
\begin{equation}
  {\bf U} = \left( \begin{array}{ccc}
            W_{11} & W_{12} & 
               q_{,1} W_{11} + q_{,2} W_{12} - \tau_0 \, F_{x_1} / F_q \\ 
            \bullet & W_{22} & 
               q_{,1} W_{12} + q_{,2} W_{22} - \tau_0 \, F_{x_2} / F_q \\
            \bullet & \bullet &
            q_{,1}^2 W_{11} + 2 q_{,1} q_{,2} W_{12} + q_{,2}^2 W_{22} -        
               \tau_0 \, (q_{,1} F_{x_1} + q_{,2} F_{x_2} + F_{x_3}) / F_q
                   \end{array} \right) \, .
  \label{eq10}
\end{equation}
Here  $q \equiv p_3$ is the vertical component of the slowness vector, $q_{,i} \equiv \partial q / \partial p_i = -F_{p_i} / F_q$, $F_{p_k} \equiv \partial F/\partial p_k$, $F_q \equiv \partial F/\partial q$ and $F_{x_k} \equiv \partial F/\partial x_k$. The matrix~(\ref{eq10}) is symmetric, so the bullets are used to denote the elements
$U_{21} = U_{12}$, $U_{31} = U_{13}$ and $U_{32} = U_{23}$.

The derivatives $F_{x_k}$ with respect to the spatial coordinates in equation~(\ref{eq10})
depend on the medium properties (i.e., heterogeneity and anisotropy) near the CMP location. Therefore the NMO-velocity surface as a whole can be reconstructed from the NMO ellipse ${\bf W}({\bf x})$, if we know the slowness vector of the zero-offset ray and local values of the elastic constants $c_{ij}({\bf x})$ near the common midpoint.

\subsection{Homogeneous media}

Here we show that the NMO-velocity surface in homogeneous
anisotropic media always represents a cylinder with the axis parallel to the zero-offset ray. The matrix ${\bf U}$ that describes this cylinder can be obtained in closed form using the Christoffel equation expressed in terms of the slowness components.

If the medium is homogeneous, the spatial derivatives $F_{x_k}$ vanish, and the matrix  ${\bf U}$ from equation~(\ref{eq10}) simplifies to
\begin{equation}
  {\bf U}^{\rm hom} = \left( \begin{array}{ccc}
            W_{11} & W_{12} & 
               q_{,1} W_{11} + q_{,2} W_{12} \\ 
            \bullet & W_{22} & 
               q_{,1} W_{12} + q_{,2} W_{22} \\
            \bullet & \bullet &
            q_{,1}^2 W_{11} + 2 q_{,1} q_{,2} W_{12} + q_{,2}^2 W_{22} 
                   \end{array} \right) .
  \label{eq11}
\end{equation}
To find the shape of the corresponding NMO-velocity surface, note that the third column of the matrix ${\bf U}^{\rm hom}$ is a linear combination of the first two columns:
\begin{equation}
  q_{,1} \, U_{k1}^{\rm hom} + q_{,2} \, U_{k2}^{\rm hom} = 
  U_{k3}^{\rm hom} \, , \qquad (k=1,2,3) \, .
 \label{eq121}
\end{equation}
As follows from equation~(\ref{eq121}),
\begin{equation}
  \det {\bf U}^{\rm hom} = 0 \, .
  \label{eq12}
\end{equation}
Since the first and the second columns are generally independent, the matrix 
${\bf U}^{\rm hom}$ has one zero eigenvalue, so the surface defined by ${\bf U}^{\rm hom}$ has to be a cylinder. For models in which the matrix ${\bf W}$ describes an ellipse in the horizontal plane, the NMO-velocity surface is an elliptical cylinder. This conclusion is valid for any pure-mode reflections in homogeneous arbitrary anisotropic media.

The axis of the NMO-velocity cylinder is parallel to the eigenvector ${\bf e} = [e_1, e_2, e_3]$ corresponding to the zero eigenvalue of the matrix ${\bf U}^{\rm hom}$. Substituting the eigenvector ${\bf e}$ into the first two rows of the 
matrix~(\ref{eq11}) yields
\begin{equation}
  \left\{ \begin{array}{l}  
          W_{11} \, \dfrac{e_1}{e_3} + W_{12} \, \dfrac{e_2}{e_3} =
             -q_{,1} W_{11} - q_{,2} W_{12} \, , \\ \\
          W_{12} \, \dfrac{e_1}{e_3} + W_{22} \, \dfrac{e_2}{e_3} =
             -q_{,1} W_{12} - q_{,2} W_{22} \, .
          \end{array} \right. 
  \label{eq13}
\end{equation}
Therefore,
\begin{equation}
  \frac{e_i}{e_3} = -q_{,i} \, , \qquad (i=1,2) \, .
  \label{eq14}
\end{equation}

The meaning of equation~(\ref{eq14}) can be explained using the expression for the components of the group-velocity vector ${\bf g}$ obtained by Grechka, Tsvankin and Cohen\cite{GrechkaTsvankinCohen1999} [their equation~(B--3)],
\begin{equation}
  \frac{g_i}{g_3} = -q_{,i} \, , \qquad (i=1,2) \, .
  \label{eq15}
\end{equation}
Comparison of equations~(\ref{eq14}) and~(\ref{eq15}) shows that ${\bf e}$ is parallel to the vector ${\bf g}$ at the CMP location. In other words, the axis of the cylinder in homogeneous media of any symmetry points in the direction of the zero-offset ray. According to the geometrical meaning of the NMO-velocity surface, this result implies that the NMO velocity on the CMP line parallel to the zero-offset ray is infinite. Indeed, if sources and receivers are placed on the (straight) zero-offset ray, the reflected rays travel along the acquisition line; consequently, the two-way reflection traveltime in CMP geometry has to be independent of offset (i.e., $V_{\rm nmo}$ goes to infinity).

The exact equation of the NMO ellipse ${\bf W}$ in anisotropic homogeneous media is given by Grechka, Tsvankin and Cohen\cite{GrechkaTsvankinCohen1999} [equation~(7)] in terms of the slowness components,   
\begin{equation}
      {\bf W} = { {p^{}_1 q^{}_{,1} + p^{}_2 q^{}_{,2} - q} \over 
              {q^{}_{,11} q^{}_{,22} - q_{,12}^2} } 
       \left(  \begin{array}{r r}   
        q^{}_{,22} & - q^{}_{,12} \\
        - q^{}_{,12} &   q^{}_{,11} 
        \end{array} \right) ,
  \label{eq16}
\end{equation}
where 
\[
      q_{,ij} = - { {F_{p_i p_j} + F_{p_i q} q_{,j} + F_{p_j q} q_{,i} + 
                    F_{qq} q_{,i} q_{,j} } \over {F_q} } \, ,
      \qquad (i,j = 1,2) \, ,
\]
$F_{p_i p_j} \equiv {\partial^2 F} / {\partial p_i} {\partial p_j}$,
$F_{p_i q} \equiv {\partial^2 F} / {\partial p_i} {\partial q}$, and
$F_{qq} \equiv {\partial^2 F} / {\partial q^2}$. Since all quantities in
equation~(\ref{eq16}) can be obtained explicitly from the Christoffel equation, equations~(\ref{eq11}) and~(\ref{eq16}) 
indicate that the whole NMO-velocity cylinder can also be constructed {\em analytically}$\,$ for a given slowness vector ${\bf p} = [p_1, p_2, q]$ of the zero-offset ray.

\subsection{NMO cylinder of the $\bm{P}$-wave in a weakly anisotropic VTI layer} 

According to equation~(\ref{eq11}), if the NMO-velocity cylinder ${\bf U}^{\rm hom}$ has been reconstructed from seismic data, it should be possible to find the derivatives $q_{,i}$ in addition to the NMO ellipse ${\bf W}$. Since for some models $q_{,i}$ may depend on medium parameters not constrained by the NMO ellipse, the NMO-velocity surface may provide valuable information for anisotropic inversion. This point is illustrated here for the $P$-wave NMO-velocity cylinder ${\bf U}^{\rm VTI}$ from a plane dipping reflector beneath a homogeneous VTI layer.

To simplify the derivation of the matrix ${\bf U}^{\rm VTI}$,
we assume that the anisotropy is weak, and the NMO velocity can be linearized in Thomsen's\cite{Thomsen1986} anisotropic coefficients $\epsilon$ and $\delta$. The $P$-wave NMO ellipse [i.e., the elements $W_{ij}$ in 
equation~(\ref{eq11}) expressed through the horizontal slowness components 
$p_1$ and $p_2$] in VTI media is fully controlled by the NMO velocity from a horizontal reflector

\begin{equation}
   V_{{\rm nmo}, P} = V_{P0} \, \sqrt{1 + 2 \, \delta}
  \label{eq17}
\end{equation}
and the anellipticity coefficient $\eta$ defined as 
\begin{equation}
   \eta \equiv \frac{\epsilon - \delta}{1 + 2 \, \delta} \, \, ,
  \label{eq18}
\end{equation}
where $V_{P0}$ is the $P$-wave vertical velocity\cite{AlkhalifahTsvankin1995, GrechkaTsvankin1998}. Hence, it is instructive to express our results in terms of $V_{{\rm nmo},P}$, $\eta$, and one of the generic Thomsen parameters (e.g., $\delta$) instead of the more conventional parameter set [$V_{P0}$, $\epsilon$, $\delta$]. 

\begin{figure}
	\centerline{\begin{tabular}{c}
			\includegraphics[width=10.5cm]{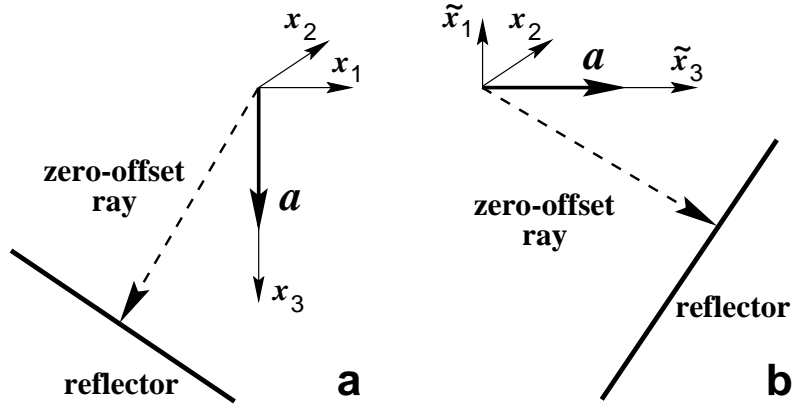} \\
	\end{tabular}}
	\caption{Dipping reflectors beneath VTI (a) and HTI (b) media. The HTI model 
		is obtained by rotating the symmetry axis ${\bf a}$ of the VTI model by 90$^\circ$ around the strike direction $x_2$.         
	}
	\label{fig02}
\end{figure}

Selecting the coordinate frame in which the reflector normal lies in the vertical plane $[x_1, x_3]$, so that the zero-offset slowness component $p_2 = 0$ (Figure~\ref{fig02}a), we use equations~(\ref{eq11}) and~(\ref{eq16}) to obtain 
\begin{eqnarray}
  && U^{\rm VTI}_{11} \equiv W^{\rm VTI}_{11} = 
    \frac{1}{V_{{\rm nmo},P}^2} - p_1^2 \, 
      \left[1 + 2 \, \eta \, (4 \, y^2 - 9 \, y + 6) \right] , 
  \label{eq11-11} \\
  && U^{\rm VTI}_{12} \equiv W^{\rm VTI}_{12} = 0 \, ,
  \label{eq11-12} \\
  && U^{\rm VTI}_{22} \equiv W^{\rm VTI}_{22} = 
    \frac{1}{V_{{\rm nmo},P}^2} - 2 \, \eta \, p_1^2 \, (2 - y) \, ,
  \label{eq11-22} \\
  && U^{\rm VTI}_{13} = 
    -\frac{p_1 \, \sqrt{1-y}}{V_{{\rm nmo},P}} \, 
    \left[1 + \delta -
          \eta \, y \, \frac{8 \, y^2 - 15 \, y + 8}{1-y} \right] ,
  \label{eq11-13} \\
  && U^{\rm VTI}_{23} = 0 \, ,
  \label{eq11-23} \\
  && U^{\rm VTI}_{33} = p_1^2 \, 
     \left[1 + 2 \, \delta - 4 \, \eta \, y \, (1 - 2 \, y) \right] \, ,
  \label{eq11-33}
\end{eqnarray}
where $y \equiv p_1^2 \, V_{{\rm nmo},P}^2$. 

Equations~(\ref{eq11-11})~--~(\ref{eq11-22}) describe the NMO ellipse ${\bf W}^{\rm VTI}$ with axes pointing in the dip and strike directions of the reflector. The dip component of the $P$-wave NMO velocity is defined by $W^{\rm VTI}_{11}$ in equation~(\ref{eq11-11})\cite{AlkhalifahTsvankin1995}, while
equation~(\ref{eq11-22}) for $W^{\rm VTI}_{22}$ gives the strike component\cite{GrechkaTsvankin1998}. Clearly, the NMO ellipse ${\bf W}^{\rm VTI}$ as a whole is governed by $V_{{\rm nmo},P}$ and $\eta$, with no dependence on $\delta$; this conclusion holds for strong anisotropy as well\cite{GrechkaTsvankin1998}.

Equations~(\ref{eq11-13})~--~(\ref{eq11-33}), specifying the additional components of ${\bf U}^{\rm VTI}$ needed to build the NMO-velocity cylinder, indicate that the $V_{\rm nmo}$ in non-horizontal directions depends on all three parameters ($V_{{\rm nmo},P}$, $\eta$ and $\delta$). This result also follows from the equation of the NMO ellipse in TI media with a horizontal symmetry axis (HTI) given in\cite{Contrerasetal1999HTI}. Note that the vertical symmetry axis ${\bf a}$ in Figure~\ref{fig02}a becomes horizontal after rotating the whole plot by 90$^\circ$ around the coordinate axis $x_2$. Hence, this rotation transforms the VTI model in Figure~\ref{fig02}a into the HTI model in Figure~\ref{fig02}b. The quantities $U^{\rm VTI}_{22}$ and $U^{\rm VTI}_{33}$ determine the NMO ellipses in both the vertical $[x_2, x_3]$-plane of the original VTI model
and the horizontal $[x_2, \tilde x_3]$-plane of the new HTI model. The results of Contreras et al.\cite{Contrerasetal1999HTI} show that the HTI ellipse ${\bf W}^{\rm HTI}$ is a function of $\delta$ as well as of $V_{{\rm nmo},P}$ and $\eta$.

Although the discussion of NMO velocities measured outside the horizontal plane may seem purely academic (unless vertical or oblique boreholes are available), 
intersections of the NMO-velocity surface with non-horizontal planes 
play an important role in Dix-type averaging of NMO velocities in heterogeneous anisotropic media. As we demonstrate below, the information contained in the matrix elements $U_{k3}$ can be extracted from the surface NMO ellipse in 
the presence of lateral heterogeneity above the reflector (such as 
intermediate dipping interfaces). It is known, though, that this is not possible for media with elliptical anisotropy\cite{DellingerMuir1988}.

\section{Dix-type formulae in heterogeneous anisotropic media}

The NMO-velocity surfaces ${\bf U}({\bf x})$ can be called ``effective'' because, just as for effective NMO velocities, they incorporate the influence of the medium properties along the whole ray path between the zero-offset reflection point and the CMP location. Here, we devise Dix-type formulae for building the effective NMO-velocity surfaces from interval (or local) surfaces in heterogeneous anisotropic media.

\subsection{General considerations}

Let us assume that the projection of the slowness vector ${\bf p}$ onto a certain plane ${\cal P}$ is preserved along the segment $L$ of the zero-offset ray. That will be the case, for example, if the ray crosses homogeneous layers separated by plane parallel interfaces ${\cal P}$. For simplicity, suppose that this segment starts at the zero-offset reflection point. To find the NMO-velocity surface ${\bf U}$ at the end of the segment $L$, it is convenient to rewrite equation~(\ref{eq02}) in the vector form,
\begin{equation}
  \tau_0 \, {\bf U}^{-1} = \frac{\partial {\bf x}}{\partial {\bf p}} \, .
  \label{eq20}
\end{equation}
Constructing the intersection of the surface ${\bf U}$ with the plane 
${\cal P}$, we obtain the NMO ellipse 
${\bf W}^{\cal P} = {\bf U} \bigcap {\cal P}$, which satisfies
\begin{equation}
  \tau_0 \, \left[ {\bf W}^{\cal P} \right]^{-1} = 
  \frac{\partial {\bf x}^{\cal P}}{\partial {\bf p}} \, 
  \Biggl|_{{\bf p}^{\cal P}}\, .
  \label{eq21}
\end{equation}
The superscript ``${\cal P}$'' in ${\bf x}^{\cal P}$ and ${\bf p}^{\cal P}$ 
emphasizes that the derivatives of the ray coordinates ${\bf x}$ are taken in
the plane ${\cal P}$ $({\bf x}^{\cal P} = {\bf x} \in {\cal P})$ with respect 
to the projection of the slowness vector ${\bf p}$ onto this plane\cite{GrechkaTsvankinCohen1999}.

Next, we divide the segment $L$ into a number of smaller
intervals $\ell$ and define the interval zero-offset traveltimes $\tau_{0,\ell}$ and 
NMO ellipses ${\bf W}^{\cal P}_{\ell}$. The interval NMO ellipses correspond to non-existent reflectors orthogonal to the slowness vector of the zero-offset ray.  
Applying equation~(\ref{eq21}) to each interval yields
\begin{equation}
  \tau_{0,\ell} \, \left[ {\bf W}^{\cal P}_{\ell} \right]^{-1} = 
  \frac{\partial {\bf x}_{\ell}^{\cal P}}{\partial {\bf p}} \, 
  \Biggl|_{{\bf p}^{\cal P}} \, .
  \label{eq22}
\end{equation}
Summing up equation~(\ref{eq22}) over the segment $L$ and taking into account that ${\bf x}^{\cal P} = \sum_\ell {\bf x}^{\cal P}_\ell$ and $\tau_0 = \sum_\ell \tau_{0,\ell}$, we find 

\begin{equation}
  \left[ {\bf W}^{\cal P} \right]^{-1} {\sum_\ell \tau_{0,\ell}} =
  \sum_\ell \tau_{0,\ell} \, \left[ {\bf W}^{\cal P}_{\ell} \right]^{-1} \, .
  \label{eq24}
\end{equation}
Equation~(\ref{eq24}) is identical to the generalized Dix formula of Grechka, Tsvankin and Cohen\cite{GrechkaTsvankinCohen1999} derived for horizontally layered media (i.e., for a horizontal plane ${\cal P}$) above a dipping reflector. Note, however, that the plane ${\cal P}$ in equation~(\ref{eq24}) can have an arbitrary orientation.

The derivation above can be repeated for the segment $L$ located anywhere on the zero-offset ray. To obtain the intersection of the NMO-velocity surface ${\bf U}$ with the plane ${\cal P}$ at the end of $L$ we have to compute $\tau_0 \, \left[ {\bf W}^{\cal P} \right]^{-1}$ at the beginning of the segment and add it to the right-hand side of equation~(\ref{eq24}). Therefore, if the projection of the slowness vector ${\bf p}^{\cal P}$ onto the plane ${\cal P}$ is preserved along the segment $L$, the contribution of this segment to the intersection of the effective NMO-velocity surface with ${\cal P}$ can be obtained using Dix-type averaging of the corresponding intersections 
${\bf W}^{\cal P}_{\ell}$ of the interval NMO-velocity surfaces ${\bf U}_{\ell}$. 

Below, we demonstrate how the effective NMO velocity in heterogeneous anisotropic media can be obtained by integrating the interval NMO ellipses along the zero-offset ray.

\subsection{Heterogeneous anisotropic media}

\begin{figure}
	\centerline{\begin{tabular}{c}
			\includegraphics[width=7.5cm]{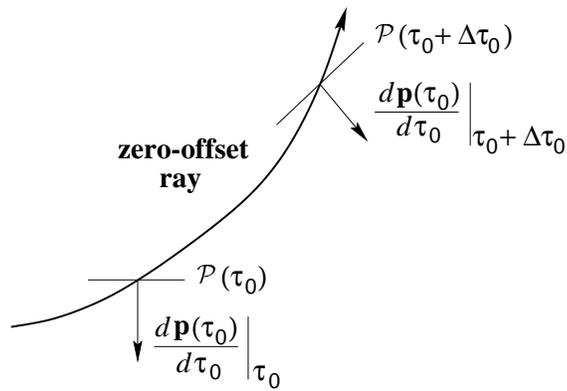} \\
	\end{tabular}}
	\caption{If the projection of the slowness vector ${\bf p}(\tau_0)$
		onto the plane ${\cal P}(\tau_0)$ (${\cal P}(\tau_0)$ is orthogonal to $d {\bf p} / d \tau_0$) is locally preserved, the intersection of the NMO-velocity surface with ${\cal P}(\tau_0)$ at the end of the ray segment corresponding to $\Delta \tau$ can be found from the Dix-type equation~(\protect\ref{eq24}). 
	}
	\label{fig04}
\end{figure}

Suppose the medium  is heterogeneous, and all components of the slowness vector vary in some fashion along the zero-offset ray. The description of rays in heterogeneous anisotropic media is given by the following system of differential equations\cite{CervenyMolotkovPsencik1977}: 
\begin{equation}
  \frac{d {\bf x}}{d \tau_0} = \frac{\partial H}{\partial {\bf p}}  
  \label{eq25}
\end{equation}
and
\begin{equation}
    \frac{d {\bf p}}{d \tau_0} = -\frac{1}{2} \, 
                               \frac{\partial H}{\partial {\bf x}} \, ,  
  \label{eq25a}
\end{equation}
where $\tau_0$ is the traveltime along the ray, and 
$H \equiv H({\bf p}, {\bf x})$ is the Hamiltonian of some particular form, which does not need to be specified here. Equation~(\ref{eq25a}) indicates that the slowness ${\bf p}$ changes in the direction $d {\bf p} / d \tau_0$ as we move along the ray. Hence, the projection of the slowness vector onto the tangent plane 
${\cal P}(\tau_0) \, \bot \, d {\bf p} / d \tau_0$ (i.e., the plane ${\cal P}(\tau_0)$ is orthogonal to the vector $d {\bf p} / d \tau_0$) is locally preserved (Figure~\ref{fig04}). Therefore,
equation~(\ref{eq24}) can be applied to the NMO ellipse   
${\bf W}^{{\cal P}(\tau_0)}(\tau_0) = {\bf U}(\tau_0) \bigcap {\cal P}(\tau_0)$
at the infinitesimal ray segment corresponding to the interval traveltime 
$\Delta \tau_0$ to produce the ellipse 
${\bf W}^{{\cal P}(\tau_0)}(\tau_0 + \Delta \tau_0)$.

To account for the fact that ${\cal P}(\tau_0 + \Delta \tau_0)$ generally
differs from ${\cal P}(\tau_0)$ (Figure~\ref{fig04}), we reconstruct
the whole NMO surface ${\bf U}(\tau_0 + \Delta \tau_0)$ 
[using the Christoffel equation] and find its intersection with the plane 
${\cal P}(\tau_0 + \Delta \tau_0)$ (see Appendices C and D):
\[
   {\bf W}^{{\cal P}(\tau_0 + \Delta \tau_0)}(\tau_0 + \Delta \tau_0) = 
   {\bf U}(\tau_0 + \Delta \tau_0) \bigcap 
   {\cal P}(\tau_0 + \Delta \tau_0) \, .
\]
 The resulting NMO ellipse ${\bf W}^{{\cal P}(\tau_0 + \Delta \tau_0)}(\tau_0 + \Delta \tau_0)$ can be continued along the next time interval, starting at 
$\tau_0 + \Delta \tau_0$. Using this formalism, the NMO surface can be built by integrating the local NMO ellipses while solving the ray-tracing equations in heterogeneous anisotropic media. A more detailed mathematical description of this procedure is given in Appendices C and D. On the whole, the results above show that it is possible to model NMO ellipses in heterogeneous anisotropic media by tracing a single (zero-offset) ray.

\subsection{Homogeneous layers separated by plane dipping interfaces}

\begin{figure}
	\centerline{\begin{tabular}{c}
			\includegraphics[width=12.0cm]{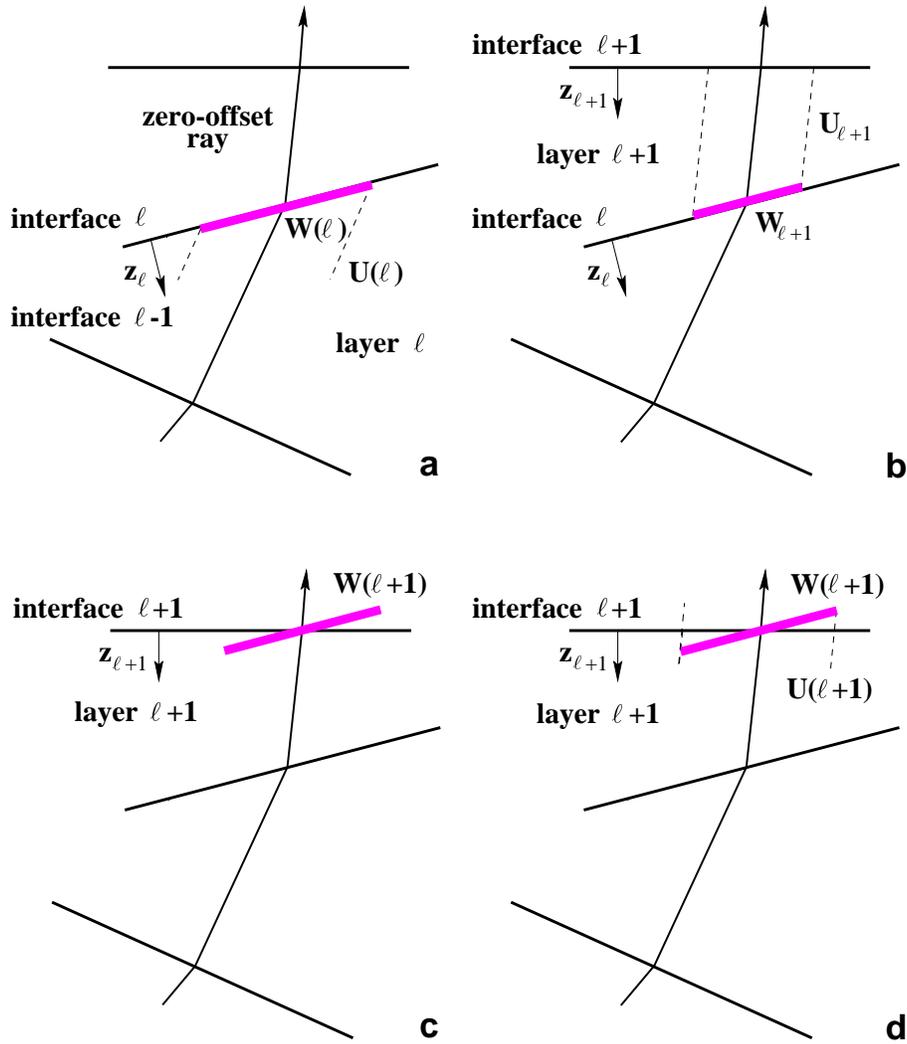} \\
	\end{tabular}}
	\caption{Dix-type averaging of the NMO-velocity cylinders over a stack of homogeneous anisotropic layers separated by plane dipping 
		interfaces. The cross-section ${\bf W}(\ell)$~(a) of the effective cylinder ${\bf U}(\ell)$ at the top of the $\ell$th layer and 
		the interval cross-section ${\bf W}_{\ell+1}$~(b) in the $\ell+1$th 
		layer are averaged using equation~(\protect\ref{eq38}) to produce 
		the cross-section ${\bf W}(\ell+1)$~(c) of the effective cylinder ${\bf U}(\ell+1)$~(d) at the surface. Note that ${\bf W}(\ell+1)$ is the intersection of ${\bf U}(\ell+1)$ with a plane parallel to the $\ell$th
		interface. The vector ${\bf z}_\ell$ is orthogonal to the $\ell$th interface.
	}
	\label{fig05}
\end{figure}

The theory of the Dix-type averaging of NMO-velocity surfaces yields relatively simple results for the practically important special case of piecewise homogeneous media composed of anisotropic layers (or blocks) separated by plane dipping 
interfaces. In such a medium, the projection of the zero-offset slowness vector onto each interface is preserved due to Snell's law (i.e., ${\bf p}_{\ell} \times {\bf z}_{\ell} = {\bf p}_{\ell+1} \times {\bf z}_{\ell}$
at the $\ell$th interface with the normal ${\bf z}_{\ell}$; see Figure~\ref{fig05}).
Therefore, the layer boundaries play the role of the planes 
${\cal P}$ that determine the intersections ${\bf W}^{\cal P}$ to be averaged by the Dix-type equation. Note that, as shown above [equation~(\ref{eq11})], NMO-velocity surfaces in piecewise homogeneous media always have a cylindrical shape.

Figure~\ref{fig05} schematically illustrates the 3-D process of constructing the NMO-velocity cylinders in layered media with dipping interfaces. Assuming that the slownesses ${\bf p}_{\ell}$ and 
traveltimes $\tau_{0,\ell}$ have already been obtained from ray tracing, the Dix-type averaging can be performed as follows:
\begin{description}
\item[Step 1.]
Using equations~(\ref{eq11}) and~(\ref{eq16}), compute the NMO-velocity cylinder ${\bf U}_1$ in the layer $\ell = 1$ immediately above the reflector; the slowness vector ${\bf p}_1$ is parallel to the reflector normal. The interval cylinder ${\bf U}_1$ is equal to the 
``effective'' cylinder ${\bf U}(1)$ in the first layer. If the layer
number $\ell > 1$, the cylinder ${\bf U}(\ell)$ (dashed lines in 
Figure~\ref{fig05}a) is obtained from the continuation procedure described here.
\item[Step 2.]
Apply equation~(\ref{eq07c}) to determine the intersection 
${\bf W}(\ell)$ (the magenta line in Figure~\ref{fig05}a) of the cylinder
${\bf U}(\ell)$ with the $\ell$th interface that has the normal ${\bf z}_\ell$.
\item[Step 3.]
Compute the interval cylinder ${\bf U}_{\ell+1}$ (dashed lines in
Figure~\ref{fig05}b) using equations~(\ref{eq11}) and~(\ref{eq16}) for the 
slowness vector ${\bf p}_{\ell+1}$. Find the intersection ${\bf W}_{\ell+1}$ 
(the magenta line in Figure~\ref{fig05}b) of the cylinder ${\bf U}_{\ell+1}$ with the 
$\ell$th interface.
\item[Step 4.]
Obtain the cross-section ${\bf W}(\ell+1)$ of the effective cylinder ${\bf U}(\ell+1)$ at the top of the $\ell+1$th layer (the magenta line in
Figure~\ref{fig05}c) from the Dix-type formula~(\ref{eq24}):
\begin{equation}
  \left[ {\bf W}(\ell+1) \right]^{-1} =
  \frac{\tau_0(\ell) \, \left[ {\bf W}(\ell) \right]^{-1} +
        \tau_{0,\ell+1} \, \left[ {\bf W}_{\ell+1} \right]^{-1} }
       {\tau_0(\ell+1)} \, ,
  \label{eq38}
\end{equation}
where $\tau_0(\ell) = \sum_{\jmath=1}^{\ell} \tau_{0,\jmath}$.
\item[Step 5.]
Reconstruct the cylinder ${\bf U}(\ell+1)$ (dashed lines in
Figure~\ref{fig05}d) using equations~(\ref{eq12c}) and~(\ref{eq11}).
\item[Step 6.]
Repeat {\bf Step 2} for the next ($\ell+1$)th layer. 
\end{description}

\renewcommand\thetable{\arabic{table}}
\begin{table}[ht]
	\def\arraystretch{1.5}
\centerline{
	\begin{tabular}{|| c | c | c | c | c | c | c | r ||} \hline \hline
		Reflector &$V_{P0}$& $\epsilon$ & $\delta$ & $\nu$ & $\beta$ & 
		$\phi_1$ & $\phi_2$~ \\ 
		depth (km) &(km/s) &       &          &       &         &      & \\ \hline \hline
		1.0  & 0.5    & 0.20       & 0.10     & 10.0  & 60.0    & 20.0 & 20.0 \\ \hline
		2.0  & 1.0    & 0.10       & 0.07     & 20.0  & 50.0    & 40.0 & 60.0 \\ \hline
		3.0  & 2.0    & 0.15       & 0.10     & 30.0  & 40.0    & 30.0 &  0.0 \\ \hline
		\hline
		\end {tabular}
	}
	\def\arraystretch{1.0} \vspace{1mm}
	\caption{Relevant parameters of the layered TI model with a tilted symmetry axis used in the numerical test. Parameters $\nu$ and $\beta$ are the tilt and azimuth of the symmetry axis (respectively); $\phi_1$ and $\phi_2$ are the dip and azimuth of the reflector (all in degrees).
	} \label{tab1}
\end{table}

\begin{figure}
	\centerline{\begin{tabular}{c}
			\includegraphics[width=9.5cm]{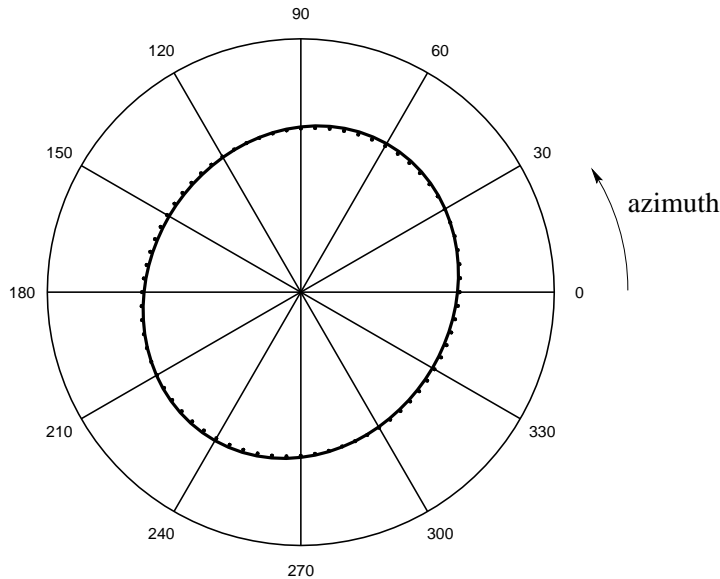} \\
	\end{tabular}}
	\caption{NMO ellipses computed for the $P$-wave reflection from the deepest
		interface in the TTI model from Table~\ref{tab1}. 
		Solid -- the ellipse calculated from the Dix-type formulae;
		dotted -- the ellipse reconstructed from the best-fit moveout velocities 
		obtained using ray-traced traveltimes in six azimuths separated by $30^\circ$.  The maximum offset $X = 3$~km is equal to the distance between the CMP and the reflector.
	}
	\label{fig06}
\end{figure}

The sequence described above makes it possible to compute NMO ellipses for layered media with plane dipping interfaces without multi-azimuth, multi-offset ray tracing. Our Dix-type averaging procedure was used by Grechka, Pech and Tsvankin\cite{GrechkaPechTsvankin2000a, GrechkaPechTsvankin2000b} to devise an efficient algorithm for $P$-wave stacking-velocity tomography in piecewise homogeneous VTI media. 

\subsection{Numerical example}

To verify the accuracy of Dix-type averaging in layered media with dipping interfaces, we computed the NMO ellipses at the horizontal surface for a 
model composed of three transversely isotropic layers with a tilted 
symmetry axis (Table 1). Figure~\ref{fig06} displays the NMO ellipses for the reflection from the bottom of the model determined from the Dix-type averaging procedure (solid)
and 3-D anisotropic ray tracing (dotted). The ellipses almost coincide, thus confirming that the Dix-type equations give an adequate description of reflection moveout on conventional-length spreads. The small difference of up 
to 1.6\% between the theoretical and ray-traced ellipses in Figure~\ref{fig06} can be attributed to the influence of nonhyperbolic moveout, which is not taken into
account by our NMO-velocity equations. However, the errors due to nonhyperbolic moveout are small (at least, for $P$-waves), when the maximum offset does not exceed roughly the distance between the CMP and the reflector. This conclusion holds for $P$-wave data in a wide variety of anisotropic models of different complexity\cite{Tsvankin2001, TsvankinThomsen1994, GrechkaTsvankin1999ORT}.

\section{Discussion}

\subsection{General results}
This work introduces the concept of NMO-velocity surfaces in 
anisotropic heterogeneous media and applies the new theory to devise Dix-type averaging procedures for effective NMO velocity. Because the NMO-velocity surface is quadratic, it depends on six generally independent elements of a symmetric matrix $U_{km}$, which include both effective quantities averaged between the reflector and CMP location and local quantities defined at the common midpoint. If the medium near the CMP is homogeneous, the NMO-velocity surface always represents a cylinder, irrespective of the complexity of the model as a whole. Other shapes that may be encountered even in isotropic media are an ellipsoid and a one-sheeted hyperboloid.

The surfaces ${\bf U}$ provide the most general description of conventional-spread normal moveout because they can be used to determine NMO velocity in any direction in 3D space. One important practical example is the NMO ellipse formed by NMO velocities plotted in all possible azimuths within a certain 
plane. Since the NMO ellipse can be viewed as a cross-section of the 
NMO-velocity surface ${\bf U}$, all properties of NMO 
ellipses examined by Grechka and Tsvankin\cite{GrechkaTsvankin1998} can be derived 
from the general expressions for ${\bf U}$ given here. Analysis of NMO-velocity surfaces also helps reveal new properties of NMO velocity, which are hidden at a less general level.

Application of these general concepts leads to  Dix-type formulae for effective normal-moveout velocity that involve averaging of specific cross-sections of the NMO-velocity surface along the zero-offset ray in heterogeneous anisotropic media. To implement this Dix-type formalism, we derived analytic expressions for computing the NMO-velocity surface and its cross-sections, as well as reconstructing the surface from a single cross-section. It should be emphasized that our averaging procedure does not require any of the slowness components of the zero-offset ray to be preserved between the reflector and the surface. (It is assumed, however, that surfaces of constant slowness, such as boundaries between layers, are locally plane at each point of the ray trajectory.) 

Although there exists an alternative way of building the NMO ellipses in heterogeneous media using dynamic ray-tracing equations\cite{GrechkaTsvankinCohen1999}, the methodology developed here is much more suitable for obtaining closed-form analytic solutions for NMO velocity; it also lends itself to geometrical interpretation. For instance, the Dix-type averaging becomes a purely analytic procedure for the important model composed of homogeneous anisotropic layers separated by plane arbitrary dipping interfaces. 

\subsection{Implications for anisotropic parameter estimation}

The main results of our Dix-type formulation which have important implications in the estimation of anisotropic parameters from reflection data can be summarized as follows: 

1. The Dix-type equations operate with the intersections ${\bf W}^{\cal P}$ of
the NMO surfaces ${\bf U}$ with generally dipping planes
${\cal P}$ determined by either dipping interfaces along the ray (Figure~\ref{fig05}) or the derivative of the slowness vector
$d {\bf p} / d \tau_0$ (Figure~\ref{fig04}).

2. The intersections ${\bf W}^{\cal P} = {\bf U} \bigcap {\cal P}$ can depend on
the anisotropic parameters that are not constrained by the NMO ellipses
in the horizontal plane. 

Thus, certain types of lateral heterogeneity may actually help in anisotropic 
inversion by tilting the planes ${\cal P}$ encountered by the zero-offset ray. 
For example, non-horizontal cross-sections of the $P$-wave NMO surface 
in VTI media depend on the individual values of the vertical velocity $V_{P0}$ 
and the anisotropic parameters $\epsilon$ and $\delta$, while the NMO ellipse 
in the horizontal plane is controlled by just their combinations $V_{{\rm nmo},P}$ 
and $\eta$ [equations~(\ref{eq11-11})~--~(\ref{eq11-33})]. As a result, the 
vertical velocity, which determines the depth scale of the model, may be 
obtained from surface reflection $P$-wave data acquired over a certain class 
of laterally heterogeneous VTI models. An example presented by Le Stunff et al.\cite{LeStunffetal1999} corroborates this conclusion for a two-layer model containing a VTI layer separated by a dipping interface from an isotropic layer.

Grechka, Pech and Tsvankin\cite{GrechkaPechTsvankin2000a, GrechkaPechTsvankin2000b} used the Dix-type equations presented here to develop algorithms for $P$-wave stacking-velocity tomography in piecewise-homogeneous VTI media. 
Their results show that the presence of irregular interfaces may also aid anisotropic 
parameter estimation in depth (e.g., using reflection tomography) by increasing the angle coverage of reflected rays. A complex subsurface structure, on the other hand, may produce trade-offs between the anisotropic velocity field and the shape of the reflector and intermediate interfaces. Understanding of the properties of NMO-velocity surfaces should help in analyzing these trade-offs and searching for practical ways to overcome them.

\section{Conclusions}

1. The pure-mode NMO velocity $V_{\rm nmo}$, treated as a function
of the direction ${\cal L}$ in 3-D space, forms a quadratic surface 
that usually is an ellipsoid, an elliptical cylinder, or a one sheeted hyperboloid. If the medium near the common midpoint is homogeneous,
the NMO-velocity surface there always has the shape of a cylinder. 

2. The NMO ellipse examined by Grechka and Tsvankin\cite{GrechkaTsvankin1998} is the intersection of the NMO-velocity surface with the horizontal plane.
 
3. The effective NMO ellipse at the surface can be obtained by Dix-type averaging of specifically oriented cross-sections of the NMO-velocity surfaces along the zero-offset ray. This formalism can be applied to any ($P$ or $S$) pure-mode reflection event.

4. The NMO-velocity surface in each anisotropic layer or block encountered by the ray  usually depends on more anisotropic parameters than does the intersection of this surface with a horizontal plane (i.e., the NMO ellipse). If the subsurface contains dipping interfaces above the reflector or other types of lateral heterogeneity, these additional parameters contribute to reflection traveltimes measured at the surface and, in some cases, can be estimated using the Dix-type formulae given here. This conclusion is valid for arbitrary anisotropy in each layer, although the parameter-estimation procedure becomes more complicated for lower symmetries. 

\section{Acknowledgments}

We are grateful to members of the A(nisotropy)-Team of the Center for Wave
Phenomena (CWP), Colorado School of Mines, for helpful discussions. Careful reviews by Joe Dellinger (BP Amoco), Ken Larner (CSM) and Ray Brown (Oklahoma Geological Survey, associate editor of Geophysics) helped to improve the manuscript.
The support for this work was provided by the members of the Consortium 
Project on Seismic Inverse Methods for Complex Structures at CWP and by the 
United States Department of Energy (Award \#DE-FG03-98ER14908).
I. Tsvankin was also supported by the Shell Faculty Career Initiation Grant.

\appendix

\section{Derivation of the NMO-velocity surface}

Here we generalize the work of Grechka and Tsvankin\cite{GrechkaTsvankin1998} on the NMO  
ellipse by developing a small-offset approximation for the squared  
reflection traveltime $t^2$ recorded on a straight common-midpoint (CMP) line  
parallel to an arbitrary unit vector ${\cal L}$. The derivation is based  
on expanding the traveltime in a Taylor series in half-offset $h$ near the  
CMP location ($h=0$). The traveltime field is assumed to be smooth  
enough for all needed derivatives to exist at zero offset.

\begin{figure}
	\centerline{\begin{tabular}{c}
			\includegraphics[width=7.5cm]{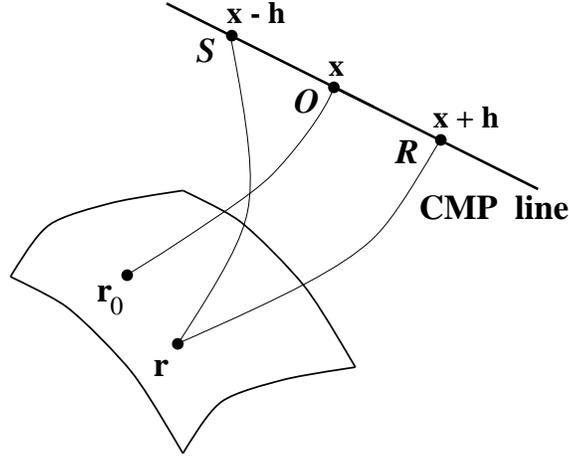} \\
	\end{tabular}}
	\caption{Reflected rays recorded on a common-midpoint line in 3D space. Point
		${\bf r}_0$ is the reflection point of the zero-offset ray originated at  
		the CMP location $O$. The ray excited at $S$ and emerging at $R$ is  
		reflected at a different point ${\bf r}$, but the reflection-point dispersal has no influence on NMO velocity.
	}
	\label{fig01a}
\end{figure}

If the coordinate of the common midpoint $O$  is denoted by ${\bf y}$ 
(Figure~\ref{fig01a}), the  
coordinates of the source $S$ and receiver $R$  are 
${\bf y - x}$ and ${\bf y + x}$, where
\begin{equation}
  {\bf x} \equiv [x_1, \, x_2, \, x_3] = h {\cal L} 
          \equiv h \, [{\cal L}_1, \, {\cal L}_2, \, {\cal L}_3] \, .  
  \label{eq01a}
\end{equation}
The pure-mode two-way reflection traveltime $t \,$ depends on the positions
of the source and receiver and on the coordinate 
${\bf r} = {\bf r}({\bf y}, {\bf x})$ of the reflection point. Summing up the 
one-way traveltimes $\tau$ corresponding to the downgoing and upgoing rays 
yields
\begin{equation}
  t({\bf y}, {\bf x}, {\bf r}) = 
  \tau({\bf y} - {\bf x}, {\bf r}) + \tau({\bf y} + {\bf x}, {\bf r}) \, .
  \label{eq02a}
\end{equation}

Since reflection point dispersal (i.e., the deviation of ${\bf r}$ from 
${\bf r}_0$ in Figure~\ref{fig01a} for $|{\bf x}| \ne 0$) does not change NMO velocity\cite{HubralKrey1980, Goldin1986}, we assume that nonzero-offset rays are reflected at point ${\bf r}_0$,
\[
  t({\bf y}, {\bf x}, {\bf r}) = t({\bf y}, {\bf x}, {\bf r}_0) = 
  \tau({\bf y} - {\bf x}, {\bf r}_0) + \tau({\bf y} + {\bf x}, {\bf r}_0) \, .
  \]
Hence, for a given reflection event and a fixed CMP location (constant ${\bf y}$ and ${\bf r}_0$), $t$ is a function of the one-way traveltime from the zero-offset reflection point,
\begin{equation}
t({\bf x}) = 
  \tau(- {\bf x}) + \tau({\bf x}) \, .
\label{eq021a}
\end{equation}

Taking into account equation~(\ref{eq01a}), the first derivative of the 
traveltime with respect to the half-offset $h$ is given by
\begin{equation}
  \frac{dt}{dh} = \sum_{k=1}^3 \frac{\partial t}{\partial x_k} {\cal L}_k \, .
  \label{eq04a}
\end{equation}
Using equation~(\ref{eq021a}), we find the derivative~(\ref{eq04a}) at zero offset as
\begin{equation}
  \frac{dt}{dh} \, \biggl|_{h=0} = 
  \sum_{k=1}^3 \biggl[ 
      -\frac{\partial \tau}{\partial x_k} 
      +\frac{\partial \tau}{\partial x_k}
               \biggl] \, {\cal L}_k = 0 \, .
  \label{eq07a}
\end{equation}

Differentiating equation~(\ref{eq04a}) again yields
\begin{equation}
  \frac{d^2 t}{d h^2} = 
  \sum_{k,m=1}^3 \frac{\partial^2 t}{\partial x_k \, \partial x_m} \, 
                 {\cal L}_k \, {\cal L}_m \, .
  \label{eq08a}
\end{equation}
Therefore, at zero offset 
\begin{equation}
  \frac{d^2 t}{d h^2} \, \biggl|_{|{\bf x}|=h=0} = 
  \sum_{k,m=1}^3 \frac{\partial^2 t}
                      {\partial x_k \, \partial x_m} \, \biggl|_{h=0}
                 {\cal L}_k \, {\cal L}_m =
  2 \sum_{k,m=1}^3 \frac{\partial^2 \tau}
                        {\partial x_k \, \partial x_m} \, \biggl|_{h=0} \, 
                 {\cal L}_k \, {\cal L}_m \, ,
  \label{eq11a}
\end{equation}

To obtain the equation for NMO velocity along the CMP line ${\cal L}$, we
expand the traveltime $t(h)$ in a Taylor series,
\begin{equation}
  t(h,{\cal L}) = t_0 +  
                  \frac{d t}{d h} \, \biggl|_{h=0} \, h +
                  \frac{d^2 t}{d h^2} \, \biggl|_{h=0} \, \, \frac{h^2}{2} + 
                  \ldots \, ,
  \label{eq12a}
\end{equation}
where $t_0 = 2\tau_0$ is the two-way zero-offset traveltime. Substituting equations~(\ref{eq07a}) and~(\ref{eq11a})  into equation~(\ref{eq12a}) leads to
\begin{equation}
  t(h,{\cal L}) = t_0 +
  h^2 \sum_{k,m=1}^3 \frac{\partial^2 \tau}
                        {\partial x_k \, \partial x_m} \, \biggl|_{h=0} \,
                 {\cal L}_k \, {\cal L}_m + \ldots 
  \label{eq13a}
\end{equation}

Squaring equation~(\ref{eq13a}) and keeping quadratic and lower-order terms with
respect to $h$ yields 
\begin{equation}
  t^2(h,{\cal L}) = t_0^2 +
  2 \, t_0 \, h^2 \sum_{k,m=1}^3 \frac{\partial^2 \tau}
                                      {\partial x_k \, \partial x_m} \, \biggl|_{h=0} \, {\cal L}_k \, {\cal L}_m \, .     
  \label{eq14a}
\end{equation}
Introducing the source-receiver offset
\begin{equation}
  X = 2 \, h \, , 
  \label{eq15a}
\end{equation}
we rewrite equation~(\ref{eq14a}) in its final form
\begin{equation}
  t^2(X,{\cal L}) = t_0^2 + 
                    ({\cal L} \, {\bf U} \, {\cal L}^{\bf T}) \, X^2 \, .
  \label{eq16a}
\end{equation}
Here the superscript ${\bf T}$ denotes transposition; the $3 \times 3$ symmetric matrix ${\bf U}$ is defined as
\begin{equation}
  U_{km} \equiv \tau_0 \, \frac{\partial^2 \tau}
                          {\partial x_k \, \partial x_m} \, \biggl|_{h=0} =
           \tau_0 \, \frac{\partial p_k}{\partial x_m} \, \biggl|_{h=0} \, ,  \qquad
           (k,m = 1,2,3) \, , 
  \label{eq17a}
\end{equation}
and
\begin{equation}
  p_k = \frac{\partial \tau}{\partial x_k}  \, , \qquad
           (k = 1,2,3) \,  
  \label{eq18a}
\end{equation}
are the components of the slowness vector ${\bf p} = [p_1, p_2, p_3]$.

Comparing equation~(\ref{eq16a}) with the conventional definition of the
normal-moveout velocity $V_{\rm nmo}({\cal L})$ on the CMP line ${\cal L}$,

\begin{equation}
  t^2(X,{\cal L}) = t_0^2 + \frac{X^2}{V_{\rm nmo}^2({\cal L})} \, ,  
  \label{eq19a}
\end{equation}
we conclude that
\begin{equation}
  \frac{1}{V_{\rm nmo}^2({\cal L})} = 
  {\cal L} \, {\bf U} \, {\cal L}^{\bf T} \, .
  \label{eq20a}
\end{equation}

\section{Constructing the NMO-velocity surface from the NMO ellipse}

Let us assume that the NMO ellipse (matrix ${\bf W}$) in the $[x_1, x_2]$-plane has been reconstructed from three or more moveout-velocity measurements in different azimuthal directions. To build the NMO-velocity surface ${\bf U}$ from ${\bf W}$ using equation~(\ref{eq07}), we need to obtain the matrix elements $U_{k3}$. This can be done by using the Christoffel equation, 
\begin{equation}
  F({\bf p}, {\bf x}) = 0 \, ,  
  \label{eq01b}
\end{equation}
where ${\bf p} = {\bf p}({\bf x})$ is the slowness vector of rays generated at the zero-offset reflection point, and $F$ explicitly depends on the spatial coordinates ${\bf x}$ because in heterogeneous media 
the elastic stiffness coefficients $c_{ij}$ vary in space.

Differentiating equation~(\ref{eq01b}) with respect to $x_1$ and $x_2$ yields
\begin{equation}
  \sum_{k=1}^3 F_{p_k} \, \frac{\partial p_k}{\partial x_j} +
               F_{x_j} = 0 \, , \qquad (j=1,2) \, ,  
  \label{eq02b}
\end{equation}
where $F_{p_k} \equiv \partial F / \partial p_k$ and 
$F_{x_j} \equiv \partial F / \partial x_j$.  The
partial derivatives of the horizontal slowness components and the zero-offset traveltime $\tau_0$ define the NMO ellipse ${\bf W}$ [see equations~(\ref{eq02}) and~(\ref{eq06})]:
\begin{equation}
  \frac{\partial p_i}{\partial x_j} = \frac{W_{ij}}{\tau_0} \, ,
  \qquad (i,j=1,2) \, .         
  \label{eq03b}
\end{equation}

Substituting equation~(\ref{eq03b}) into equation~(\ref{eq02b}) and solving 
for $\partial p_3 / \partial x_j$, we find
\begin{equation}
  U_{j3} \equiv \tau_0 \, \frac{\partial p_3}{\partial x_j} =
  - \frac{F_{p_1} \, W_{1j} + F_{p_2} \, W_{2j} + \tau_0 \, F_{x_j}}
         {F_{p_3}} \, , \qquad (j=1,2) \, .
  \label{eq04b}
\end{equation}
Differentiating the Christoffel equation~(\ref{eq01b}) 
with respect to $x_3$ and taking into account that $\partial p_j / \partial x_3 = \partial p_3 / \partial x_j$
because of the symmetry of ${\bf U}$ [equations~(\ref{eq02})
and~(\ref{eq07})] leads to the following expression for $U_{33}$:  
\begin{equation}
  U_{33} \equiv \tau_0 \, \frac{\partial p_3}{\partial x_3} =
  - \frac{F_{p_1} \, U_{13} + F_{p_2} \, U_{23} + \tau_0 \, F_{x_3}}
         {F_{p_3}} \, .
  \label{eq05b}
\end{equation}

At a fixed spatial location ${\bf x}$ the Christoffel equation
can be treated as a relationship between the vertical 
slowness component $p_3 \equiv q$ and the horizontal slownesses $p_1$ 
and $p_2$. Implicit differentiation of the Christoffel equation then gives\cite{GrechkaTsvankinCohen1999}
\begin{equation}
   q_{,j} = - \frac{F_{p_j}}{F_q} \, , \qquad (j=1,2) \, ,
  \label{eq06b}
\end{equation}
where 
$q_{,j} \equiv \partial q / \partial p_j \equiv \partial p_3 / \partial p_j$
and $F_q \equiv \partial F / \partial q \equiv \partial F / \partial p_3$.
Using equation~(\ref{eq06b}), we rewrite equations~(\ref{eq04b}) 
and~(\ref{eq05b}) in the form
\begin{equation}
  U_{j3} = q_{,1} W_{1j} + q_{,2} W_{2j} - 
  \tau_0 \, \frac{F_{x_j}}{F_q} \, , \qquad (j=1,2) \, 
  \label{eq07b}
\end{equation}
and
\begin{equation}
  U_{33} = q_{,1} U_{13} + q_{,2} U_{23} -
  \tau_0 \, \frac{F_{x_3}}{F_q} \, . 
  \label{eq08b}
\end{equation}

Substituting equation~(\ref{eq07b}) into~(\ref{eq08b}), we obtain the final
expression for $U_{33}$ and the NMO-velocity surface as a whole: 
\begin{equation}
  {\bf U} = \left( \begin{array}{ccc}
            W_{11} & W_{12} & 
               q_{,1} W_{11} + q_{,2} W_{12} - \tau_0 \, F_{x_1} / F_q \\
            \bullet & W_{22} &
               q_{,1} W_{12} + q_{,2} W_{22} - \tau_0 \, F_{x_2} / F_q \\
            \bullet & \bullet &
            q_{,1}^2 W_{11} + 2 q_{,1} q_{,2} W_{12} + q_{,2}^2 W_{22} -        
               \tau_0 \, (q_{,1} F_{x_1} + q_{,2} F_{x_2} + F_{x_3}) / F_q
                   \end{array} \right) \, ,
  \label{eq09b}
\end{equation}
where bullets in the low left-hand corner of the matrix denote the elements 
$U_{21} = U_{12}$, $U_{31} = U_{13}$ and $U_{32} = U_{23}$.

\section{Dix-type averaging in heterogeneous anisotropic media}

Here we give a detailed description of the Dix-type procedure for building the NMO-velocity surfaces in heterogeneous
anisotropic media. Suppose we would like to use a known surface 
${\bf U}(\tau_0)$ at the zero-offset traveltime $\tau_0$ to construct
 the surface ${\bf U}(\tau_0 + \Delta \tau_0)$ at  the time 
$\tau_0 + \Delta \tau_0$, where $\Delta \tau_0$ is an infinitesimal interval. We assume that the projection of the slowness vector ${\bf p}(\tau_0)$ onto the plane ${\cal P}(\tau_0) \, \bot \, d {\bf p} / d \tau_0$ is locally preserved 
over the ray segment corresponding to $\Delta \tau_0$ (Figure~\ref{fig04}).  If we find the intersection of the NMO-velocity surface ${\bf U}(\tau_0)$ with the plane ${\cal P}(\tau_0)$,
\[
   {\bf W}^{{\cal P}(\tau_0)}(\tau_0) = {\bf U}(\tau_0) \bigcap 
                                        {\cal P}(\tau_0) \, ,
\]
the intersection at the time $\tau_0 +\Delta \tau_0$ is given by the Dix-type equation~(\ref{eq24}):
\begin{equation}
  \left[ {\bf W}^{{\cal P}(\tau_0)}(\tau_0 + \Delta \tau_0) \right]^{-1} =
  \frac{\tau_0 \, \left[ {\bf W}^{{\cal P}(\tau_0)}(\tau_0) \right]^{-1} + 
        \Delta \tau_0 \, \left[ {\bf W}^{{\cal P}(\tau_0)}  
           \bigl|_{[\tau_0, \, \tau_0 + \Delta \tau_0]}\right]^{-1} }
       {\tau_0 + \Delta \tau_0} \, .
  \label{eq26}
\end{equation}
Here 
\[
   {\bf W}^{{\cal P}(\tau_0)} \bigl|_{[\tau_0, \, \tau_0 + \Delta \tau_0]} = 
   {\bf U} \bigl|_{[\tau_0, \, \tau_0 + \Delta \tau_0]} \bigcap 
   {\cal P}(\tau_0) \,  
\]
is the intersection of the {\it local}$\,$ NMO-velocity surface  
${\bf U} \bigl|_{[\tau_0, \, \tau_0 + \Delta \tau_0]}$ with the plane ${\cal P}(\tau_0)$. 
The local surface ${\bf U} \bigl|_{[\tau_0, \, \tau_0 + \Delta \tau_0]}$ can be 
computed using equations~(\ref{eq10}) and~(\ref{eq16}).

The plane ${\cal P}(\tau_0 + \Delta \tau_0) \, \bot \, 
d {\bf p} / d \tau_0 \bigl|_{\tau_0 + \Delta \tau_0}$ at the traveltime 
$\tau_0 + \Delta \tau_0$ generally differs from the plane 
${\cal P}(\tau_0)$, as shown in Figure~\ref{fig04}. In order to account for
the rotation of the plane ${\cal P}$ along the ray, we reconstruct the NMO 
surface ${\bf U}(\tau_0 + \Delta \tau_0)$ from its cross-section
${\bf W}^{{\cal P}(\tau_0)}(\tau_0 + \Delta \tau_0)$ using 
equations~(\ref{eq09c}) and~(\ref{eq12c}) (see Appendix~D) and find the intersection of ${\bf U}(\tau_0 + \Delta \tau_0)$ with the plane 
${\cal P}(\tau_0 + \Delta \tau_0)$:
\[
   {\bf W}^{{\cal P}(\tau_0 + \Delta \tau_0)}(\tau_0 + \Delta \tau_0) = 
   {\bf U}(\tau_0 + \Delta \tau_0) \bigcap 
   {\cal P}(\tau_0 + \Delta \tau_0) \, .
\]
Since the slowness components are locally preserved in the plane 
${\cal P}(\tau_0 + \Delta \tau_0)$, we can continue the cross-section
${\bf W}^{{\cal P}(\tau_0 + \Delta \tau_0)}(\tau_0 + \Delta \tau_0)$
along the next ray segment using equation~(\ref{eq26}) applied to the 
time interval $[\tau_0 + \Delta \tau_0, \, \tau_0 + 2 \Delta \tau_0]$.
Thus, NMO-velocity surfaces in heterogeneous anisotropic media can be computed simultaneously with integrating ray equations~(\ref{eq25}). 

In summary, continuation of the NMO-velocity surface along the zero-offset ray involves the following steps:
\begin{description}
\item[Step 1.]
Construct the intersection ${\bf W}^{{\cal P}(\tau_0)}(\tau_0)$ 
[equation~(\ref{eq07c}) below] of the given NMO-velocity surface ${\bf U}(\tau_0)$ 
with the plane ${\cal P}(\tau_0)$ orthogonal to the vector $d {\bf p} / d \tau_0$, which is 
specified by the second equation~(\ref{eq25}). 
\item[Step 2.]
Continue the cross-section of the NMO-velocity surface over the time interval $\Delta \tau_0$; that is, compute ${\bf W}^{{\cal P}(\tau_0)}(\tau_0 + \Delta \tau_0)$ from ${\bf W}^{{\cal P}(\tau_0)}(\tau_0)$ using  
equation~(\ref{eq26}).
\item[Step 3.]
Reconstruct the surface ${\bf U}(\tau_0 + \Delta \tau_0)$ 
[equations~(\ref{eq09c}) and~(\ref{eq12c}) below] from its
cross-section ${\bf W}^{{\cal P}(\tau_0)}(\tau_0 + \Delta \tau_0)$.
\item[Step 4.]
Repeat {\bf Step 1} for the surface ${\bf U}(\tau_0 + \Delta \tau_0)$.
\end{description}

\section{Operations with cross-sections of NMO-velocity surfaces}

In this appendix, we show how to compute the intersection 
${\bf W}^{\cal P}$ of the NMO-velocity surface ${\bf U}$ with an arbitrary 
plane 
${\cal P}$ and reconstruct the matrix ${\bf U}$ from its given cross-section 
${\bf W}^{\cal P}$. Let us denote by ${\bf z}$ the unit vector in the direction 
$d {\bf p} / d \tau_0$ [see equation~(\ref{eq25})] normal to the plane 
${\cal P}$. The vector ${\bf z}$ can be specified by two spherical angles
$\phi_1$ and $\phi_2$:
\begin{equation}
  {\bf z} = [\sin \phi_1 \cos \phi_2, \,
             \sin \phi_1 \sin \phi_2, \, \cos \phi_1] \, .
  \label{eq01c}
\end{equation}
It is straightforward to verify that the unit vectors
\begin{equation}
  {\bf b}^{(1)} = [\cos \phi_1 \cos \phi_2, \,
                   \cos \phi_1 \sin \phi_2, \, -\sin \phi_1] \,  
  \label{eq02c}
\end{equation}
and
\begin{equation}
  {\bf b}^{(2)} = [-\sin \phi_2, \, \cos \phi_2, \, 0] \,
  \label{eq03c}
\end{equation}
are both orthogonal to ${\bf z}$ and, therefore, lie in the plane
${\cal P} \perp {\bf z}$. Thus, any vector ${\bf b}$ in ${\cal P}$ is given by
\begin{equation}
  {\bf b} = {\bf b}^{(1)} \cos \alpha + {\bf b}^{(2)} \sin \alpha \, ,
  \label{eq04c}
\end{equation}
where $\alpha$ is the azimuth (within ${\cal P}$) with respect to 
${\bf b}^{(1)}$.

The NMO velocity [equation~(\ref{eq01})] within the plane ${\cal P}$,   
\begin{equation}
  \frac{1}{V_{\rm nmo}^2({\bf b})} = 
     {\bf b} \, {\bf U} \, {\bf b}^{\bf T} \, ,
  \label{eq05c}
\end{equation}
can be viewed as the intersection of the NMO surface ${\bf U}$ with the plane 
${\cal P}$. Substituting equations~(\ref{eq02c}) -- (\ref{eq04c}) 
into~(\ref{eq05c}) yields
\begin{equation}
  \frac{1}{V_{\rm nmo}^2(\alpha)} \Biggl|_{\cal P} = 
  W_{11}^{\cal P} \cos^2 \alpha + 
  2 \, W_{12}^{\cal P} \sin \alpha \cos \alpha +
  W_{22}^{\cal P} \sin^2 \alpha \, ,
  \label{eq06c}
\end{equation}
where
\begin{equation}
  W_{ij}^{\cal P} = \sum_{k,m=1}^3 B_{km, \,ij} \, U_{km} \, , 
  \qquad (i,j=1,2) \, ,
  \label{eq07c}
\end{equation}
and
\begin{equation}
  B_{km, \,ij} = \frac{1}{2} \, \left( b_k^{(i)} b_m^{(j)} + 
                                       b_k^{(j)} b_m^{(i)} \right) \, ,
  \qquad (i,j=1,2; ~~ k,m=1,2,3) \, .
  \label{eq08c}
\end{equation}
Equations~(\ref{eq01c}) -- (\ref{eq03c}), (\ref{eq07c}), and~(\ref{eq08c}) 
define the matrix ${\bf W}^{\cal P}$ that describes the intersection
(i.e., the NMO ellipse) of the NMO surface ${\bf U}$ with the plane ${\cal P}$
with the unit normal ${\bf z}$.

Next, we show how to reconstruct the whole NMO surface ${\bf U}$ from its
cross-section ${\bf W}^{\cal P}$. This procedure is based on 
equation~(\ref{eq09b}), which can be written in the form
\begin{equation}
  {\bf U} = \left( \begin{array}{ccc}
            W_{11} & W_{12} & 
               q_{,1} W_{11} + q_{,2} W_{12} + A_1 \\ 
            \bullet & W_{22} & 
               q_{,1} W_{12} + q_{,2} W_{22} + A_2 \\ 
            \bullet & \bullet &
            q_{,1}^2 W_{11} + 2 q_{,1} q_{,2} W_{12} + q_{,2}^2 W_{22} + A_3 
                   \end{array} \right) \, ,
  \label{eq09c}
\end{equation}
where ${\bf W}$ is the NMO ellipse in the horizontal plane 
$[x_1, x_2]$, and the quantities $A_i$ are given by  
\[
  A_i = -\tau_0 \, \frac{F_{x_i}}{F_q} \, , \qquad (i=1,2) \, ,
\]
\[
  A_3 = -\tau_0 \, \frac{q_{,1} F_{x_1} + q_{,2} F_{x_2} + F_{x_3}}{F_q} \, .
\]
The derivatives of $F$ [equation~(\ref{eq01b})], which also determine
the derivatives $q_{,i}$ and $q_{,ij}$  of the vertical slowness component with respect to the horizontal slownesses 
[equation~(\ref{eq06b}) and the second equation~(\ref{eq16})], are evaluated 
at point ${\bf x}$ of the zero-offset ray specified by the one-way 
traveltime $\tau_0$.

It is evident from equation~(\ref{eq09c}) that in order to compute ${\bf U}$
we need to know the matrix ${\bf W}$ because all other quantities are
obtained by differentiating the Christoffel equation~(\ref{eq01b}). 
Substituting equation~(\ref{eq09c}) into~(\ref{eq07c}) leads to three linear
equations relating the matrices ${\bf W}$ and
${\bf W}^{\cal P}$:
\begin{equation}
  W_{ij}^{\cal P} = C_{ij} + \sum_{i',j'=1}^2 D_{i'j', \,ij} \, W_{i'j'} \, , 
  \qquad (i,j=1,2) \, .
  \label{eq10c}
\end{equation}
Here
\[
  C_{ij} = 2 A_1 B_{13, \,ij} + 2 A_2 B_{23, \,ij} + A_3 B_{33, \,ij} \,
\]
and
\[
  D_{i'j', \,ij} = B_{i'j', \,ij} + 
                   q_{,i'} B_{j'3, \,ij} + q_{,j'} B_{i'3, \,ij} +
                   q_{,i'} q_{,j'} B_{33, \,ij} \, , 
  \qquad (i',j',i,j=1,2) \, .
\]
To emphasize the fact that equations~(\ref{eq10c}) represent a system 
of linear equations for the unknown elements $W_{i'j'}$, we replace the pairs of 
indexes $\{ij\}$ and $\{i'j'\}$ by a single index using the following convention: $\{11\} \rightarrow 1$, $\{12\} \rightarrow 2$, and $\{22\} \rightarrow 3$. 
Then, equations~(\ref{eq10c}) can be rewritten in a more conventional form,
\begin{equation}
  \sum_{k'=1}^3 E_{k'k} \, W_{k'} = C_k - W_k^{\cal P}   \, , 
  \qquad (k=1,2,3) \, .
  \label{eq11c}
\end{equation}
where $E_{k'k}$ are the elements of the $3 \times 3$ matrix 
\[
  {\bf E} = \left( \begin{array}{c c c}
	D_{11, \,11} & 2 \, D_{12, \,11} & D_{22, \,11} \\
    D_{11, \,12} & 2 \, D_{12, \,12} & D_{22, \,12} \\
     D_{11, \,22} & 2 \, D_{12, \,22} & D_{22, \,22} \\  
     \end{array} \right) .
\]
The linear system~(\ref{eq11c}) can be solved for the matrix ${\bf W}$ using standard techniques: 
\begin{equation}
  {\bf W} = {\bf E}^{-1} \left( {\bf W}^{\cal P} - {\bf C} \right) \, .
  \label{eq12c}
\end{equation}
Thus, equations~(\ref{eq09c}) and~(\ref{eq12c}), supplemented with the Christoffel equation~(\ref{eq01b}), make it possible to reconstruct the NMO-velocity surface ${\bf U}$ from its intersection ${\bf W}^{\cal P}$ with the plane 
${\cal P}$.  

%\bibliography{/Users/VG/Dropbox/VG-File-Exchange/DCP/text/references/refs-vg-DCP} % Mac
%\bibliography{C:/Users/erf/Dropbox/VG-File-Exchange/DCP/text/references/refs-vg-DCP} % PC 
\bibliography{refs-vg-DCP}

\newcommand{\SortBy}[1]{}
\begin{thebibliography}{23}
\expandafter\ifx\csname natexlab\endcsname\relax\def\natexlab#1{#1}\fi
\expandafter\ifx\csname bibnamefont\endcsname\relax
  \def\bibnamefont#1{#1}\fi
\expandafter\ifx\csname bibfnamefont\endcsname\relax
  \def\bibfnamefont#1{#1}\fi
\expandafter\ifx\csname citenamefont\endcsname\relax
  \def\citenamefont#1{#1}\fi
\expandafter\ifx\csname url\endcsname\relax
  \def\url#1{\texttt{#1}}\fi
\expandafter\ifx\csname urlprefix\endcsname\relax\def\urlprefix{URL }\fi
\providecommand{\bibinfo}[2]{#2}
\providecommand{\eprint}[2][]{\url{#2}}

\bibitem[{\citenamefont{Tsvankin}(2001)}]{Tsvankin2001}
\bibinfo{author}{\bibfnamefont{I.}~\bibnamefont{Tsvankin}},
  \emph{\bibinfo{title}{Seismic signatures and analysis of reflection data in
  anisotropic media}}, vol.~\bibinfo{volume}{29} of
  \emph{\bibinfo{series}{Handbook of Geophysical Exploration (3rd ed., 2012,
  SEG)}} (\bibinfo{publisher}{Elsevier}, \bibinfo{year}{2001}).

\bibitem[{\citenamefont{Grechka and Tsvankin}(1998)}]{GrechkaTsvankin1998}
\bibinfo{author}{\bibfnamefont{V.}~\bibnamefont{Grechka}} \bibnamefont{and}
  \bibinfo{author}{\bibfnamefont{I.}~\bibnamefont{Tsvankin}},
  \bibinfo{journal}{Geophysics} \textbf{\bibinfo{volume}{63}},
  \bibinfo{pages}{no.\ 3, 1079} (\bibinfo{year}{1998}).

\bibitem[{\citenamefont{Levin}(1971)}]{Levin1971}
\bibinfo{author}{\bibfnamefont{F.~K.} \bibnamefont{Levin}},
  \bibinfo{journal}{Geophy\-sics} \textbf{\bibinfo{volume}{36}},
  \bibinfo{pages}{510} (\bibinfo{year}{1971}).

\bibitem[{\citenamefont{Shah}(1973)}]{Shah1973}
\bibinfo{author}{\bibfnamefont{P.~M.} \bibnamefont{Shah}},
  \bibinfo{journal}{Geophy\-sics} \textbf{\bibinfo{volume}{38}},
  \bibinfo{pages}{812} (\bibinfo{year}{1973}).

\bibitem[{\citenamefont{Grechka
  et~al.}(1999{\natexlab{a}})\citenamefont{Grechka, Theophanis, and
  Tsvankin}}]{GrechkaTheophanisTsvankin1999}
\bibinfo{author}{\bibfnamefont{V.}~\bibnamefont{Grechka}},
  \bibinfo{author}{\bibfnamefont{S.}~\bibnamefont{Theophanis}},
  \bibnamefont{and} \bibinfo{author}{\bibfnamefont{I.}~\bibnamefont{Tsvankin}},
  \bibinfo{journal}{Geophy\-sics} \textbf{\bibinfo{volume}{64}},
  \bibinfo{pages}{no.\ 1, 146} (\bibinfo{year}{1999}{\natexlab{a}}).

\bibitem[{\citenamefont{Corrigan et~al.}(1996)\citenamefont{Corrigan, Withers,
  Darnall, and Skopinski}}]{Corriganetal1996}
\bibinfo{author}{\bibfnamefont{D.}~\bibnamefont{Corrigan}},
  \bibinfo{author}{\bibfnamefont{R.}~\bibnamefont{Withers}},
  \bibinfo{author}{\bibfnamefont{J.}~\bibnamefont{Darnall}}, \bibnamefont{and}
  \bibinfo{author}{\bibfnamefont{T.}~\bibnamefont{Skopinski}},
  \bibinfo{journal}{66th Annual International Meeting, SEG, Expanded Abstracts}
  pp. \bibinfo{pages}{1834--1837} (\bibinfo{year}{1996}).

\bibitem[{\citenamefont{Grechka and
  Tsvankin}(1999{\natexlab{a}})}]{GrechkaTsvankin1999}
\bibinfo{author}{\bibfnamefont{V.}~\bibnamefont{Grechka}} \bibnamefont{and}
  \bibinfo{author}{\bibfnamefont{I.}~\bibnamefont{Tsvankin}},
  \bibinfo{journal}{Geophysics} \textbf{\bibinfo{volume}{64}},
  \bibinfo{pages}{no.\ 4, 1202} (\bibinfo{year}{1999}{\natexlab{a}}).

\bibitem[{\citenamefont{Grechka
  et~al.}(1999{\natexlab{b}})\citenamefont{Grechka, Tsvankin, and
  Cohen}}]{GrechkaTsvankinCohen1999}
\bibinfo{author}{\bibfnamefont{V.}~\bibnamefont{Grechka}},
  \bibinfo{author}{\bibfnamefont{I.}~\bibnamefont{Tsvankin}}, \bibnamefont{and}
  \bibinfo{author}{\bibfnamefont{J.~K.} \bibnamefont{Cohen}},
  \bibinfo{journal}{Geophysical Prospecting} \textbf{\bibinfo{volume}{47}},
  \bibinfo{pages}{no.\ 2, 117} (\bibinfo{year}{1999}{\natexlab{b}}).

\bibitem[{\citenamefont{Dix}(1955)}]{Dix1955}
\bibinfo{author}{\bibfnamefont{C.~H.} \bibnamefont{Dix}},
  \bibinfo{journal}{Geophysics} \textbf{\bibinfo{volume}{20}},
  \bibinfo{pages}{68} (\bibinfo{year}{1955}).

\bibitem[{\citenamefont{Hubral and Krey}(1980)}]{HubralKrey1980}
\bibinfo{author}{\bibfnamefont{P.}~\bibnamefont{Hubral}} \bibnamefont{and}
  \bibinfo{author}{\bibfnamefont{T.}~\bibnamefont{Krey}},
  \emph{\bibinfo{title}{Interval velocities from seismic reflection time
  measurements}} (\bibinfo{publisher}{SEG}, \bibinfo{year}{1980}).

\bibitem[{\citenamefont{Grechka and
  Tsvankin}(1999{\natexlab{b}})}]{GrechkaTsvankin1999ORT}
\bibinfo{author}{\bibfnamefont{V.}~\bibnamefont{Grechka}} \bibnamefont{and}
  \bibinfo{author}{\bibfnamefont{I.}~\bibnamefont{Tsvankin}},
  \bibinfo{journal}{Geophy\-sics} \textbf{\bibinfo{volume}{64}},
  \bibinfo{pages}{820} (\bibinfo{year}{1999}{\natexlab{b}}).

\bibitem[{\citenamefont{Contreras et~al.}(1999)\citenamefont{Contreras,
  Grechka, and Tsvankin}}]{Contrerasetal1999HTI}
\bibinfo{author}{\bibfnamefont{P.}~\bibnamefont{Contreras}},
  \bibinfo{author}{\bibfnamefont{V.}~\bibnamefont{Grechka}}, \bibnamefont{and}
  \bibinfo{author}{\bibfnamefont{I.}~\bibnamefont{Tsvankin}},
  \bibinfo{journal}{Geo\-phy\-sics} \textbf{\bibinfo{volume}{64}},
  \bibinfo{pages}{1219} (\bibinfo{year}{1999}).

\bibitem[{\citenamefont{Grechka and Tsvankin}(2000)}]{GrechkaTsvankin2000TTI}
\bibinfo{author}{\bibfnamefont{V.}~\bibnamefont{Grechka}} \bibnamefont{and}
  \bibinfo{author}{\bibfnamefont{I.}~\bibnamefont{Tsvankin}},
  \bibinfo{journal}{Geo\-phy\-sics} \textbf{\bibinfo{volume}{65}},
  \bibinfo{pages}{232} (\bibinfo{year}{2000}).

\bibitem[{\citenamefont{Slotnick}(1959)}]{Slotnick1959}
\bibinfo{author}{\bibfnamefont{M.~M.} \bibnamefont{Slotnick}},
  \emph{\bibinfo{title}{Lessons in seismic computing}}
  (\bibinfo{publisher}{SEG}, \bibinfo{year}{1959}).

\bibitem[{\citenamefont{Thomsen}(1986)}]{Thomsen1986}
\bibinfo{author}{\bibfnamefont{L.}~\bibnamefont{Thomsen}},
  \bibinfo{journal}{Geophysics} \textbf{\bibinfo{volume}{51}},
  \bibinfo{pages}{1954} (\bibinfo{year}{1986}),
  \urlprefix\url{10.1190/1.1442051}.

\bibitem[{\citenamefont{Alkhalifah and
  Tsvankin}(1995)}]{AlkhalifahTsvankin1995}
\bibinfo{author}{\bibfnamefont{T.}~\bibnamefont{Alkhalifah}} \bibnamefont{and}
  \bibinfo{author}{\bibfnamefont{I.}~\bibnamefont{Tsvankin}},
  \bibinfo{journal}{Geophysics} \textbf{\bibinfo{volume}{60}},
  \bibinfo{pages}{no.\ 5, 1550} (\bibinfo{year}{1995}).

\bibitem[{\citenamefont{Dellinger and Muir}(1988)}]{DellingerMuir1988}
\bibinfo{author}{\bibfnamefont{J.}~\bibnamefont{Dellinger}} \bibnamefont{and}
  \bibinfo{author}{\bibfnamefont{F.}~\bibnamefont{Muir}},
  \bibinfo{journal}{Geo\-phy\-sics} \textbf{\bibinfo{volume}{53}},
  \bibinfo{pages}{1616} (\bibinfo{year}{1988}).

\bibitem[{\citenamefont{{\SortBy{Cerveny}}\v{C}erven\'{y}
  et~al.}(1977)\citenamefont{{\SortBy{Cerveny}}\v{C}erven\'{y}, Molotkov, and
  P\v{s}en\v{c}\'{i}k}}]{CervenyMolotkovPsencik1977}
\bibinfo{author}{\bibfnamefont{V.}~\bibnamefont{{\SortBy{Cerveny}}\v{C}erven\'{y}}},
  \bibinfo{author}{\bibfnamefont{I.~A.} \bibnamefont{Molotkov}},
  \bibnamefont{and}
  \bibinfo{author}{\bibfnamefont{I.}~\bibnamefont{P\v{s}en\v{c}\'{i}k}},
  \emph{\bibinfo{title}{Ray method in seismology}}
  (\bibinfo{publisher}{University of Karlova}, \bibinfo{year}{1977}).

\bibitem[{\citenamefont{Grechka
  et~al.}(2000{\natexlab{a}})\citenamefont{Grechka, Pech, and
  Tsvankin}}]{GrechkaPechTsvankin2000a}
\bibinfo{author}{\bibfnamefont{V.}~\bibnamefont{Grechka}},
  \bibinfo{author}{\bibfnamefont{A.}~\bibnamefont{Pech}}, \bibnamefont{and}
  \bibinfo{author}{\bibfnamefont{I.}~\bibnamefont{Tsvankin}},
  \bibinfo{journal}{70th Annual International Meeting, SEG, Expanded Abstracts}
  pp. \bibinfo{pages}{2225--2228} (\bibinfo{year}{2000}{\natexlab{a}}).

\bibitem[{\citenamefont{Grechka
  et~al.}(2000{\natexlab{b}})\citenamefont{Grechka, Pech, and
  Tsvankin}}]{GrechkaPechTsvankin2000b}
\bibinfo{author}{\bibfnamefont{V.}~\bibnamefont{Grechka}},
  \bibinfo{author}{\bibfnamefont{A.}~\bibnamefont{Pech}}, \bibnamefont{and}
  \bibinfo{author}{\bibfnamefont{I.}~\bibnamefont{Tsvankin}},
  \bibinfo{journal}{70th Annual International Meeting, SEG, Expanded Abstracts}
  pp. \bibinfo{pages}{2229--2232} (\bibinfo{year}{2000}{\natexlab{b}}).

\bibitem[{\citenamefont{Tsvankin and Thomsen}(1994)}]{TsvankinThomsen1994}
\bibinfo{author}{\bibfnamefont{I.}~\bibnamefont{Tsvankin}} \bibnamefont{and}
  \bibinfo{author}{\bibfnamefont{L.}~\bibnamefont{Thomsen}},
  \bibinfo{journal}{Geophysics} \textbf{\bibinfo{volume}{59}},
  \bibinfo{pages}{no.\ 8, 1290} (\bibinfo{year}{1994}).

\bibitem[{\citenamefont{Stunff et~al.}(1999)\citenamefont{Stunff, Grechka, and
  Tsvankin}}]{LeStunffetal1999}
\bibinfo{author}{\bibfnamefont{Y.~L.} \bibnamefont{Stunff}},
  \bibinfo{author}{\bibfnamefont{V.}~\bibnamefont{Grechka}}, \bibnamefont{and}
  \bibinfo{author}{\bibfnamefont{I.}~\bibnamefont{Tsvankin}},
  \bibinfo{journal}{69th Annual International Meeting, SEG, Expanded Abstracts}
  pp. \bibinfo{pages}{1604--1607} (\bibinfo{year}{1999}).

\bibitem[{\citenamefont{Goldin}(1986)}]{Goldin1986}
\bibinfo{author}{\bibfnamefont{S.~V.} \bibnamefont{Goldin}},
  \emph{\bibinfo{title}{Seismic traveltime inversion}}
  (\bibinfo{publisher}{SEG}, \bibinfo{year}{1986}).

\end{thebibliography}

\end{document}